# Optical Switching in Tb/Co-Multilayer Based Nanoscale Magnetic Tunnel Junctions

Sucheta Mondal,[†] Debanjan Polley,[†,‡] Akshay Pattabi,[†,§] Jyotirmoy Chatterjee,[†] David Salomoni,[⊥] Luis Aviles-Felix,[⊥] Aurélien Olivier,[⊥] Miguel Rubio-Roy,[⊥] Bernard Diény,[⊥] Liliana Daniela Buda Prejbeanu,[⊥] Ricardo Sousa,[⊥] Ioan Lucian Prejbeanu,[⊥] and Jeffrey Bokor[†,‡,*]

[†]Department of Electrical Engineering and Computer Sciences, University of California, Berkeley, California 94720, United States.

[‡]Materials Sciences Division, Lawrence Berkeley National Laboratory, Berkeley, California 94720, United States.

[§]Department of Engineering, University of San Francisco, San Francisco, California 94117, United States

[⊥]Spintec, Université Grenoble Alpes, CNRS, CEA, Grenoble INP, IRIG-SPINTEC, 38000 Grenoble, France.

[*]jbokor@berkeley.edu

ABSTRACT: Magnetic tunnel junctions (MTJs) are elementary units of magnetic memory devices. For high-speed and low-power data storage and processing applications, fast reversal by an ultrashort laser pulse is extremely important. We demonstrate optical switching of Tb/Co-multilayer-based nanoscale MTJs by combining optical writing and electrical read-out methods. A 90 fs-long laser pulse switches the magnetization of the storage layer (SL). The change in magnetoresistance between the SL and a reference layer (RL) is probed electrically across the tunnel barrier. Single-shot switching is demonstrated by varying the cell diameter from 300 nm to 20 nm. The anisotropy, magnetostatic coupling, and switching probability exhibit cell-size dependence. By suitable association of laser fluence and magnetic field, successive commutation between high-resistance and low-resistance states is achieved. The switching dynamics in a continuous film is probed with the magneto-optical Kerr effect technique. Our experimental findings provide strong support for the growing interest in ultrafast spintronic devices.

In magneto-optical recording technology, ferrimagnetic rare earth (RE)-transition metal (TM) alloys and multilayers (MLs) are seeing expanding interest due to their tunable perpendicular magnetic anisotropy (PMA) and ability to exhibit all-optical switching (AOS) phenomena.[1,2] Binary and ternary RE-TM alloys are identified as suitable candidates for designing prototypes of several spintronic devices, such as, spin-transfer-torque devices,[3,4] current-induced switching devices,[5] multilevel switching applications,[6] and perpendicular magnetic tunnel junctions (p-MTJs)[7,8] etc. Within a p-MTJ, the data can be stored and processed by monitoring the change in the tunneling magnetoresistance (TMR) which depends on the relative orientation between the magnetization of the storage layer (SL) and of the reference layer (RL). Successful AOS operation of a micron-sized p-MTJ cell consisting of GdFeCo alloy based storage layer was demonstrated by J. Y. Chen *et al*.[7] Similar successive switching operation was recently reported by L. Wang *et al*. in Co/Gd bilayer based MTJ cell down to 3 µm diameter with laser pulses as short as ~100 fs.[8] However, these materials are inherently magnetically soft and impose serious limitations to stabilize the perpendicular magnetization direction at low dimensions. While downscaling to a sub-micron-sized cell, the magnetization tries to align parallel to the sample plane due to the competition between shape anisotropy and magnetocrystalline anisotropy. Further research in this direction suggests that Tb-based MLs with comparatively high magnetic anisotropy can be suitable candidates to construct the SL within a p-MTJ.[9–11] Tb-based alloys exhibit helicity dependent and helicity-independent AOS.[12] A [Tb/Co]$_N$-based stack exchange-coupled to another CoFeB layer was engineered to construct the SL of a MTJ[9] having the PMA preserved after annealing at 250°C. The feasibility of single-shot AOS was also demonstrated, using laser pulses of 60 fs and 6 ps width, by switching the magnetization of the SL alone.[10] These [Tb/Co]$_N$-based SL were integrated within the patterned p-MTJs cells having another CoFeB layer as the RL and achieved TMR value of 38%. A Tb-based p-MTJ has been shown to preserve its PMA for the cell-size down to 50 nm.

In this paper, we report single-shot switching operation in [Tb/Co]$_N$-based p-MTJ cells with cell-diameter systematically varied from 300 nm to 20 nm. Here the optically-switchable CoFeB/Ta/[Tb/Co]$_N$ layer (SL) and the soft CoFeB layer (RL) are separated by a highly-resistive MgO tunnel barrier with high-quality interfaces. The R×A values lie in the range of 65 Ωµm$^2$ to 420 Ωµm$^2$ for these MTJs. Switching with a laser pulse from 'high-resistance' state to 'low-resistance' state ('High' and 'Low' states from here on) where magnetization of the RL and SL are in antiparallel and in parallel alignment respectively, is demonstrated by reading the magnetoresistance ($R$) signal electrically. The switching probability is estimated to be 46% in this experiment for a 300 nm-diameter cell. Magnetic coupling between the layers during optical excitation influences the switching. We demonstrated that the rate of switching between 'High' and 'Low' states for our devices can reach ~100% by setting the soft CoFeB layer (RL) in a favorable orientation with a weak magnetic field. This also helps in stabilization of the SL after being optically switched. The magnetization dynamics is studied by time-resolved magneto-optical Kerr effect (TRMOKE) technique in a continuous film with similar stack configurations.

The complete structure of the patterned p-MTJ cell is: Ta(3.0)/Pt(5.0)/Ta(5.0)/**CoFeB(1.1)**/MgO(2.3)/**CoFeB(1.3)/Ta(0.2)/[Tb(0.9)/Co(1.2)]$_5$**/Ta(2.0)/Pt(2.0), the thickness values are given in nanometers. We have highlighted the layer-specifications which denote the RL, below the MgO layer, and the SL, above the MgO layer. The schematic of the device is shown in Figure 1a. The MTJ cells with diameter: 300 nm, 200 nm, 100 nm, 80 nm, 50 nm, and 20 nm are patterned with ITO electrodes. Details of the sample fabrication method and experimental techniques can be found in the Supporting Information. A full $R$-$H$ loop obtained for a 300 nm-sized cell with a magnetic field applied out-of-plane (OOP) is shown in Figure 1b. The saturation fields for RL and SL ($H_R$ and $H_S$) are ~150 Oe and ~1500 Oe respectively for this device. The TMR is ≳ 13% and a clear contrast of signal level exists between the 'High' state and the 'Low' state even in the presence of thermal fluctuations across the electrodes. The minor $R$-$H$ loops shown in Figure 1c reveal a loop shift of ~30-50 Oe ascribed to the dipolar coupling between the magnetic layers within the cell. Estimation of the dipolar coupling from the minor $R$-$H$ loops with different MTJ cell-diameter are presented in the Supporting Information. The OOP hysteresis loop obtained from the depth-resolved static MOKE measurement[13] on the continuous film shows no considerable coupling between the SL and RL though. supporting the fact that the loop shift observed in the patterned devices is of magnetostatic origin in contrast to Neel coupling or direct coupling through the MgO barrier. The detailed mapping of coercivity, resistance, and TMR values obtained from the current-in-plane-tunneling (CIPT) measurement on these devices can be found elsewhere.[10]

To demonstrate the single shot switching in a 300 nm-MTJ cell, the following sequence is conducted:

First, we applied a -2000 Oe magnetic field ($H_a$) to saturate both the RL and the SL ($|H_a|>|H_S|>|H_R|$) in the downward direction as shown in Figure 2a. This ensures a uniform initial magnetization configuration for the entire p-MTJ cell. (i) The magnetization of the RL is flipped in the upward direction by a +390 Oe field ($|H_S|>>|H_a|>|H_R|$) while the SL remains unchanged. This 'High' state is read by passing a small current and measuring the voltage across the junction. (ii) Then the field is removed, and the magnetization of the SL is switched by a single laser pulse (pulse-width: 90 fs, fluence: 55 mJ/cm$^2$). (iii) This 'Low' state is read in presence of $H_a$ = +390 Oe. (iv) Again, the field is removed, and device is exposed to the next laser pulse. Occasionally, the laser induces reorientation of both the SL and RL magnetization. Thus, the magnetization of RL is flipped upwards at step (i) again. The steps (i)-(iv), depicted in Figure 2a are repeated in sequence to reproduce the results. The variation of the resistance is plotted in Figure 2b. Based on this protocol the rate of successful switching of this device has been estimated to be ~ 46%. Here the switching is considered as 'successful' if the magnetization of the SL flips by 180° and the resistance alternates between the two extreme 'High' and 'Low' levels. We have also observed 'partial switching' and 'no-switching' events here.

Similar results have been obtained for a 100 nm-MTJ cell. The reasons for unsuccessful switching seem to be two-fold: (a) the optical switching of SL might be incomplete, (b) the switching of SL is followed by a reorientation of RL. As the single shot pulses generated from the laser carry slightly different peak energies in successive events, the change of effective fluence in turn can affect the switching of the SL. Thus, we have simultaneously recorded the energy of each pulse ($E$) with a photodiode (see Figure 2b). However, we have not found any correlation between the fluctuations (<30%) observed for the incident pump energy and the switching events. All the measurements have been performed with relatively higher pump fluence to avoid any ambiguity (i.e., $E$ is kept higher than the switching threshold but lower than the damage threshold of the devices). RL experiences the effect of a small but finite dipolar field ($H_d$ ~30 Oe) from the SL as evidenced from the shift in the minor *R-H* loops (see Supporting Information). The reading is performed in step (i) and (iii). The value of resistance changes from 'High' to 'Low' state only when the SL is optically switched. On the other hand, 'Low' state recorded at step (iii) is changed to 'High' state as recorded in step (i) indicates two possible scenarios: either RL is fixed and only SL is switched or SL and RL both are switched at step (iv). Different spin configurations observed during the switching events are depicted in the Supporting Information. We conclude that the SL switches in both directions at ultrafast time scales. While reading the state electrically in real time, the 'Low' to 'High' state switching shows lower switching probability. It seems that the parallel configuration between SL and RL is energetically favored. This can be easily understood from the previously reported behavior of magnetostatically coupled layers within the perpendicularly magnetized nanopillars[14,15] and across the arrays of in-plane (IP) nanomagnets.[16,17] In our experiment, the observed shift corresponds to an extension of the low resistance plateau for all the minor *R-H* loops. This is strong evidence of the fact that considerable amount of coupling is being experienced by the RL in 'parallel' configuration which is energetically favored. This in turn hinders the switching of SL from 'Low' state to 'High' state without reorienting the RL. Consequently, switching of the SL from one parallel configuration ends up with another parallel configuration and the resistance remains unchanged. It is pertinent to mention that a conventional p-MTJ is constructed of CoFeB free layer (soft magnetic SL) and synthetic antiferromagnet (SAF) coupled to another CoFeB pinned layer (hard magnetic RL). However, the reported coercivity of the SAF layer is comparable to CoFeB/Ta/[Tb/Co]$_N$.[8] This may cause difficulty in distinguishing the reversal from the two electrodes in the magnetoresistance measurement. Thus, our p-MTJ devices rely on the performance of RL consisting of the CoFeB layer alone. Moreover, it is also known that in response to the laser pulse, the coercivity of the magnetic material alters due to the heating. Heat assisted magnetic recording and thermomagnetic switching are common examples of this effect.[18–21] We cannot rule out the presence of laser induced transient modifications within the cell, such as, change in local magnetic moment, magnetic susceptibility, and other dynamical parameters which may occur in femtosecond to nanosecond time scales.[22]

To improve the device performance, we show that the direction of RL magnetization can be set by application of a weak magnetic field during the optical excitation as well as during the electrical read out. We have designed the experiment in the following manner (see Figure 3a). (v) The magnetization of the RL is set by +390 Oe field ($|H_S|>>|H_a|>|H_R|$) in the upward direction to obtain a 'High' state. (vi) Then in the presence of $H_a$, the device is excited by a laser pulse and is switched to a 'Low' state. (vii) Then magnetization of the RL is flipped by a -390 Oe field in the downward direction and the 'High' state is recorded. (viii) In the presence of $H_a$, the device is excited by the next laser pulse resulting in a 'Low' state again. These steps (v)-(viii) are repeated sequentially to reproduce the switching events. The change in resistance is plotted in Figure 3b. We have estimated the switching probability to be >95%. The field sequence and variation in pulse energy is shown here for a one-to-one comparison. The reliability of the switching is greatly improved following this second procedure, the assistance of the magnetic field being an important element.

Figure 4a shows that $H_R$ obtained from the minor $R$-$H$ loops (see Supporting Information), decreases with the cell diameter. To further underpin the relation between $H_R$ and the switching probability, we have conducted a series of experiments with MTJ cells of different diameters following the same field sequence as (v)-(viii) with $H_a$ = 390 Oe, 190 Oe, 160 Oe, and 50 Oe. The laser fluence is kept same for all the measurements. These experiments reveal that when the pinning field is larger than $H_R$ of each device, the switching probability is higher than 50% and approaches 100% with larger field values. On the other hand, for a field comparable to, or less than $H_R$, the probabilities reduce to almost zero with the read-write sequence followed in the experiment.

To better understand the magnetization dynamics during the laser excitation, time-resolved measurements on the continuous film have been performed by using the single-color pump-probe based TRMOKE technique.[13] The femtosecond pump pulse (fluence ~30 mJ/cm$^2$) is incident obliquely on the sample whereas the probe beam (fluence ~1 mJ/cm$^2$) is focused normally inside the pump spot to collect the polar MOKE signal. The repetition rate of the laser is about 252 kHz. Sufficiently large OOP magnetic field ($H_a$>$H_S$) is applied to ensure uniform magnetization in the probed region saturating both the RL and SL. The MOKE signal is mainly obtained from the Co layers within the Co/Tb ML. After the pump pulse, the effective magnetization is quenched and crosses zero in less than 2 ps (Figure 5a). The time-resolved signal obtained for 300 ps time window shows that the magnetization recovers up to 80% of the equilibrium value at this range due to presence of bias magnetic field (Figure 5b). Higher fluence causes heating and affects the local static and dynamic properties of the probe region.

In case of the MTJ cell, the thick ITO electrodes may absorb partially the incident fluence and thus the effective energy needed for switching might be higher than in the continuous film[24]. It is worth mentioning here that we have not observed a prominent signature of toggle switching from the Kerr microscopy image with single shot laser pulses on the film. We do not rule out the possibility that the final thickness of the individual magnetic and nonmagnetic layers in the stack has varied slightly from the targeted values. The requirement of relatively high fluence for switching both the MTJ cell and reference film may indicate a high sensitivity to the composition of the Co/Tb multilayer affecting the rate of switching. Applying external field to set the RL in desired orientation (for example, application of -$H_R$ will pin the RL in downward direction) did not help to observe switching from a 'Low' to 'High' state with optical pulses. We believe that in this case, even if the SL is switched at the ultrafast timescales, after the excitation it aligns back to the 'parallel' configuration due to combined effect of external magnetic field and dipolar coupling in longer time scale.[23] We observe no change in the resistance in real time. Thus, this effect could not be verified from the time-resolved experiment. The dynamics of an individual MTJ cell is difficult to obtain from the pump-probe measurement due to diffraction limited spot sizes of the laser beam. During large scale fabrication of the devices spatial nonuniformity in the magnetic and nonmagnetic layers could have introduced slight variations in the transport properties of the p-MTJs. This can result in changes in coercivity, saturation field and TMR values obtained for the cells even with similar diameters. However, this uncertainty does not influence our experimental findings. It is important to mention that previously in thermally assisted magnetic random access memory (TA-MRAM) devices, field writing combined with heat pulse was used to achieve write selectivity by lowering the magnetic pinning energy of the SL.[25] In the present experiment the SL switches with an optical pulse without an external magnetic field. However, the rate of commutation between 'High' to 'Low' state is improved by application of a pinning field to the RL.

In conclusion the downscaling, while retaining the desired functionalities, is one of the long-term goals for constructing magnetic memory devices. We have experimentally demonstrated optical switching operation in Tb/Co-based nanoscale MTJs having nominal diameters down to 20 nm. Initially a 'High' state is assigned to the MTJ cell by orientating the magnetization of the RL opposite to the SL with an external magnetic field. Then, the field is removed, and a 'Low' state is achieved by switching the magnetization of the SL optically. These steps are repeated sequentially to estimate the rate of switching. As the CoFeB/Ta/[Tb/Co]$_N$ layer has larger anisotropy than the reference CoFeB layer and the size of the MTJ enters nanoscale regime, magnetostatic coupling and magnetic anisotropy play pivotal role in controlling the equilibrium state and hence the rate of switching. Moreover, the performance of the device is highly sensitive to the composition of the Co/Tb multilayer. By using suitable combination of laser fluence and magnetic field uninterrupted switching is achieved. We have reproduced the 'High' to 'Low' state switching by application of a single-shot laser pulse while being assisted by a weak magnetic field which pins the magnetization of the RL with relatively low anisotropy in the desired orientation. The rate of switching varies with the strength of the magnetic field and the size of nanoscale MTJs. We established that a pinning field larger than the saturation field of the RL is necessary to achieve ~100% switching probability in these devices. The time scale of zero-crossing (< 2 ps) is investigated by means of TRMOKE measurement for a continuous film having similar specifications as the devices. We believe that engineering the stack with optimized layer thickness and fabricating a strongly anisotropic SAF layer adjacent to the RL, will further improve the device performance. We expect our findings to trigger further research unraveling more functionalities of these Tb/Co-based pMTJs.

ASSOCIATED CONTENT

Supporting Information

Sample fabrication and experimental methods; static magnetic characterization of the MTJ cells and the reference film; different magnetization configurations during the optical switching of the MTJ cell in absence of any external magnetic field.

AUTHOR INFORMATION

Corresponding Author

E-mail: jbokor@berkeley.edu

Notes

The authors declare no competing financial interest.

ACKNOWLEDGEMENT

This work was supported by ASCENT (one of the SRC/DARPA supported centers within the JUMP initiative); the Director, Office of Science, Office of Basic Energy Sciences, Materials Sciences and Engineering Division, of the U.S. Department of Energy under Contract No. DE-AC02-05-CH11231 within the Nonequilibrium Magnetic Materials Program (MSMAG); and the Berkeley Emerging Technology Research (BETR) Center. This work was also partially supported by FET-Open Grant Agreement No. 713481 (SPICE). D.S. has received funding from the European Union's Horizon 2020 research and innovation programme under Marie Skłodowska-Curie grant agreement No 861300 (COMRAD).

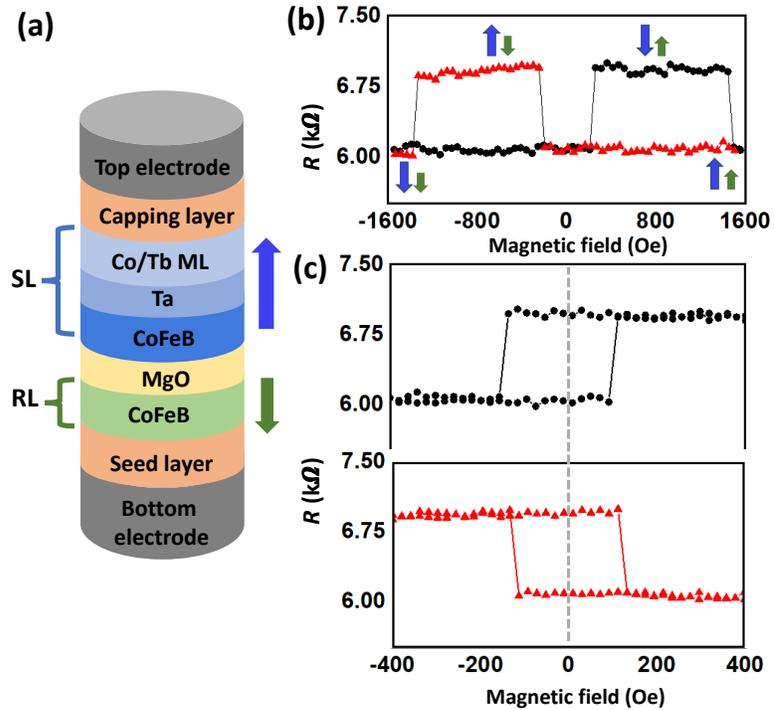

**Figure 1**: (a) Schematic of the MTJ cell. The blue and green arrows indicate the effective magnetization of SL and RL. The full stack configuration is as follows:
Ta(3.0)/Pt(5.0)/Ta(5.0)/**CoFeB(1.1)**/MgO(2.3)/**CoFeB(1.3)/Ta(0.2)/[Tb(0.9)/Co(1.2)]$_5$**/Ta(2.0)/Pt(2.0). (b) The full- and (c) half-*R-H* loops presented for an MTJ cell with 300 nm-diameter. The relative orientation of the magnetization in SL and RL at different field values are shown inside the figure. The dotted line indicates *H* = 0 Oe position.

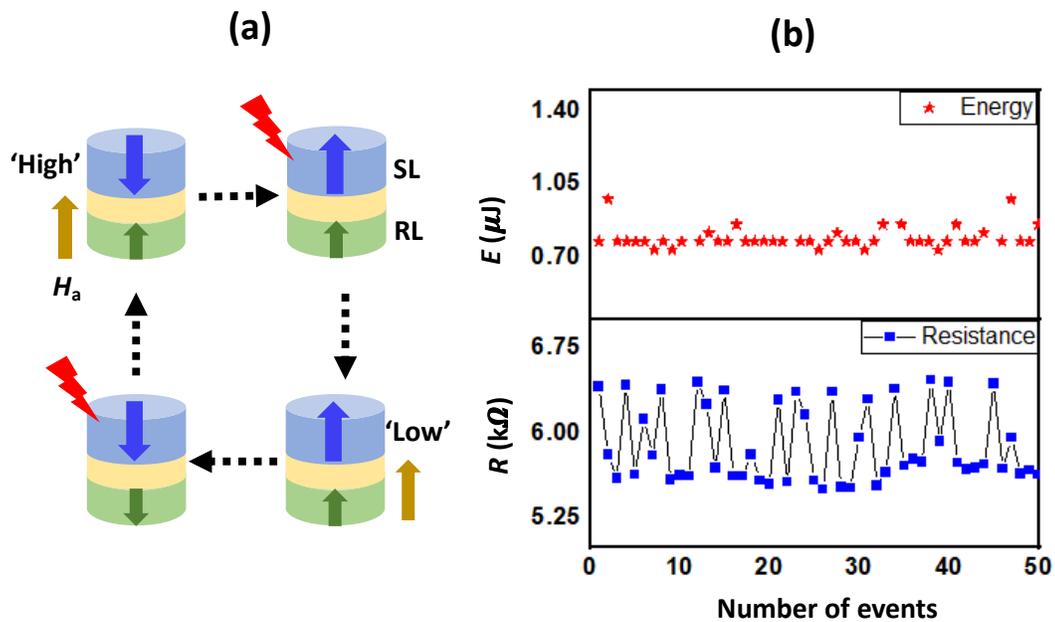

**Figure 2**: (a) For an MTJ cell of 300 nm diameter, the sequence of reading the magnetic state electrically and switching the state optically (in absence of a magnetic field) is depicted. 'High' and 'Low' stand for the high-resistance and low-resistance states respectively. The drawings represent the final configuration at each stage of the write/read process. (b) The change in magnetoresistance value (R) was recorded after the approach of each optical pulse with fluence >55 mJ/cm2. The energy of the pulse (E) was also recorded simultaneously.

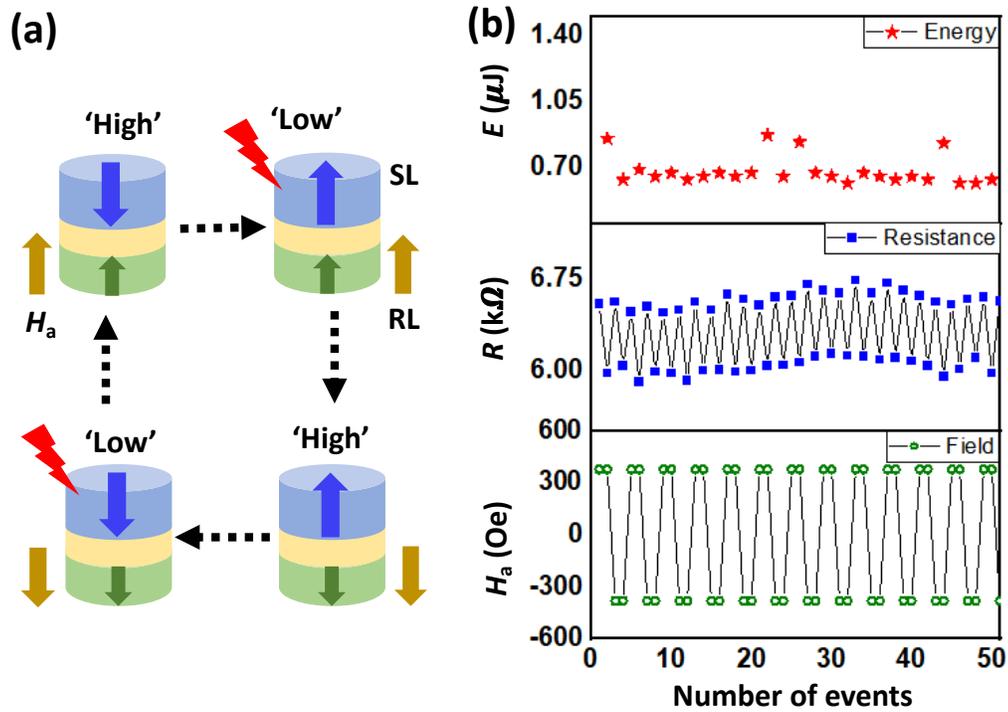

**Figure 3**: (a) For an MTJ cell of 300 nm diameter, the sequence of reading the magnetic state electrically and switching the state optically (in absence of a magnetic field) is depicted. 'High' and 'Low' stand for high-resistance and low-resistance state respectively. The drawings represent the final configuration at each stage of the write/read process. (c) The change in magnetoresistance value (R) recorded after the approach of each optical pulse with fluence >50 mJ/cm2. The energy of the pulse (E) was also recorded simultaneously. The magnetic field sequence is plotted as a reference with the number of events.

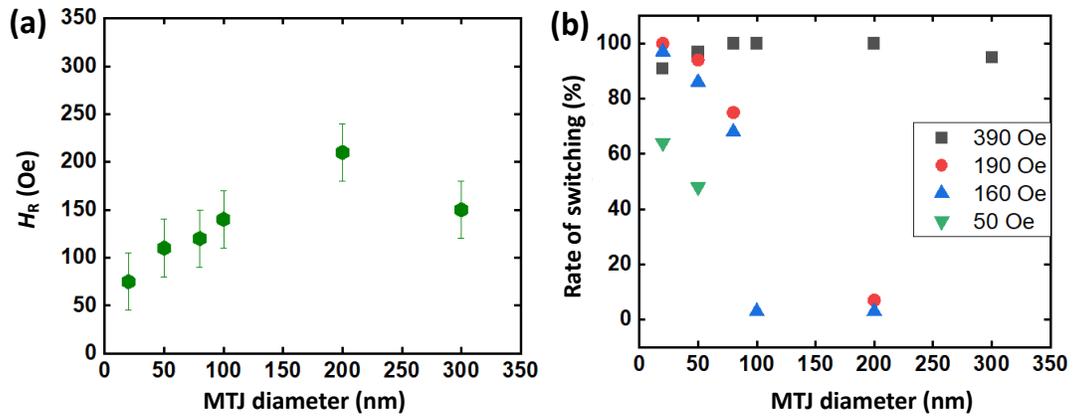

**Figure 4**: (a) The variation of saturation field (HR) for RL with the diameter of MTJ cell. (b) The variation switching rate with the diameter of MTJ cell obtained from the read-write sequence is depicted in Figure 3a. The four different values of applied magnetic field (Ha) are indicated inside the figure.

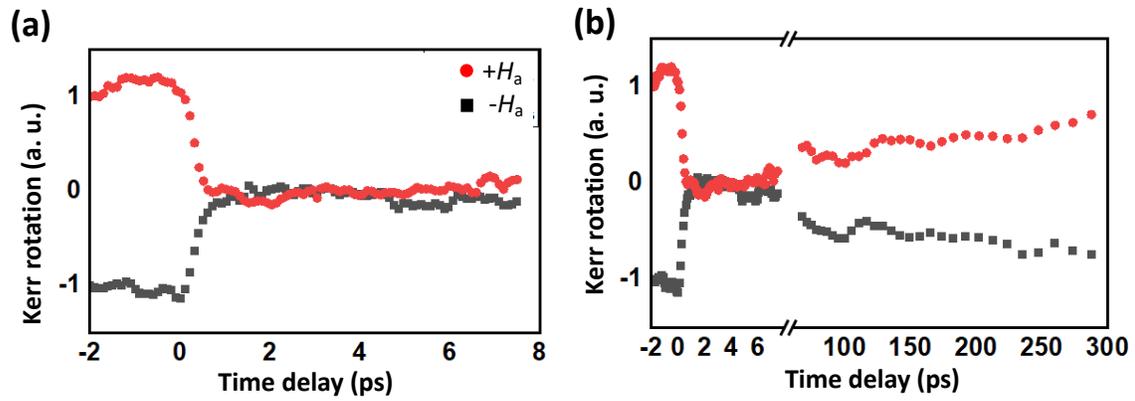

**Figure 5**: (a) The time-resolved Kerr signal obtained for the continuous film at two opposite polarities of the magnetic field, $H_a$ = 2000 Oe. The data is obtained for the range 2.0 ps in the negative delay to 7.5 ps with 100 fs temporal resolution. The applied pump-fluence and probe-fluence are >30 mJ/cm$^2$ and >1 mJ/cm$^2$ respectively. (b) The time-resolved Kerr signal data obtained up to 300 ps.

## SUPPORTING INFORMATION

**Sample fabrication and experimental methods**

The samples are grown on thermally oxidized single crystal Si (100) wafers by DC magnetron sputtering using an argon pressure of 2 mbar and a base pressure of $10^{-8}$ mbar. The structure of the magnetic layers within the MTJ is:

**CoFeB(1.1)/MgO/ CoFeB(1.3)/Ta(0.2)/[Tb(0.9)/Co(1.2)]$_5$**

The 1.1 nm-thick CoFeB reference layer (RL) is grown on the Ta(3.0)/Pt(5.0)/Ta(5.0) seed layer. Here, CoFeB(1.3)/Ta(0.2)/[Tb(0.9)/Co(1.2)]$_5$ is the optically-addressable storage layer (SL) grown on top of the MgO layer. The square brackets contain the bilayer structure that is repeated five times and thicknesses are in nanometer. Ta(2.0)/Pt(2.0) bilayer is used as capping layer to protect the magnetic stack. The MTJ cells with diameter: 300 nm, 200 nm, 100 nm, 80 nm, 50 nm, and 20 nm are patterned with 150-nm-thick ITO electrodes on top. Metallic pads of Cr/Al are fabricated adjacent to the ITO electrodes to facilitate connection with the external circuitry. The high-quality ITO deposition is essential for an efficient laser-pulse access and a reliable electrical detection. The full MTJ stack is annealed at 250° C. The annealing and the high-quality growth of MgO as a tunnel barrier, are responsible for the large tunneling magnetoresistance (TMR) achieved in these devices. Detailed description on the transport properties for the MTJs can be found elsewhere.[10] Another reference film was grown separately, to reproduce the patterned sample properties. The full stack configuration of the film is as follows:

Ta(3.0)/Pt(5.0)/Ta(5.0)/**CoFeB(1.1)**/MgO(2.3)/**CoFeB(1.3)/Ta(0.2)/[Tb(0.9)/Co(1.2)]$_5$**/Ta(2.0)/ Pt(2.0). Samples were made to have the same properties. Thickness differences might be artifacts from deposition rate drifts.

The *R-H* loops are measured in the nanoscale devices by contacting tungsten probe tips (5 micron in diameter) on the top and bottom electrodes. A current source Keithley 2400 is used to apply current, in the order of μA, and measure voltage, in the order of mV, in those devices. The value of the minimum magnetoresistance (*R*) scales with the cell-diameter. For example, the values are 6 kΩ and 220 kΩ respectively for the 300 nm-cell and the 20 nm-cell.

We used a custom-build magneto-optical microscope[13] in polar Kerr configuration to test the switching using optical pulses, wavelength ~800 nm, pulse width ~90 fs, and fluence >50 mJ/cm$^2$, sourced from a pulsed-amplified-laser system capable of generating laser pulses on a single-shot basis. Optical characterization of static and dynamics properties of the MTJ cell was not conducted due to diffraction limited spot size of the probe beam in the Kerr microscope setup.

The magnetization dynamics in the reference film is obtained by using the single-color pump-probe based time resolved magneto-optical Kerr effect (TRMOKE) technique. The femtosecond pump pulse, fluence ~30 mJ/cm$^2$, is incident obliquely on the sample whereas the probe beam, fluence ~1 mJ/cm$^2$, is focused to a spot size ~3 μm, normally inside the pump ~100 μm spot to collect the polar MOKE signal. The repetition rate of the laser is about 252 kHz. The linear polarization of the incident probe beam is modulated at 50 kHz frequency supplied from the controller of a photo elastic modulator (PEM). After being reflected from the sample surface, the beam is fed to an analyzer and then enters a photo detector. Using a phase sensitive detection method, the change in Kerr rotation is recorded in terms of voltage in the lock-in amplifier. In this set up the MOKE signal obtained from the film is mainly from the Co-based magnetic layer contribution.

We characterized the static magnetic properties of the SL and RL separately from the depth-sensitive MOKE measurement in the same experimental setup. A quarter-wave plate is placed in the probe path between the polarizer and the PEM to establish the depth-sensitive magnetic probing. By tuning the angle of the wave plate a balance between the Kerr rotation and Kerr ellipticity arising from different magnetic materials is achieved. MOKE signal from a particular layer gets suppressed and the information about rest of stack is recorded. Thus, we have extracted magnetic properties of the SL and RL separately from the reference film.

**Static magnetic characterization of the MTJ cells and the reference film**

Figure S1a shows the minor *R-H* loops obtained for the electrical measurement for MTJ cells with different diameters. With the decrease in cell diameter, the coercivity and saturation field of the devices decreases non-monotonically as evidenced from the *R-H* loops. The amount of dipolar coupling estimated from the shift of the loops is plotted in Figure S1b. This signifies that the RL and SL is coupled within the cell by the magnetic flux generated from the unsaturated spins residing at the outer surface of the cylindrical cell (Figure S1c). All the loop shifts from parallel alignment in 'Low' resistance state towards 'High' state. This is strong evidence of the fact that considerable amount of coupling is being experienced by the RL in 'parallel' configuration. This in turn hinders the switching of SL from 'Low' to 'High' state without reorienting the RL. The SL made of CoFeB(1.3)/Ta(0.2)/[Tb(0.9)/Co(1.2)]$_5$, has finite magnetic moment. The effective magnetization of Tb/Co is not exactly compensated. This introduces a finite vertical coupling with the CoFeB. As the size of the cell decreases the perpendicular magnetic anisotropy (PMA) of the layers reduces. The easy axis of the uniaxial anisotropy tries to rotate in plane and the dipolar coupling mediated by the peripheral spins between the soft- and the hard magnetic layers becomes relatively prominent. This behavior can be easily understood from the similar type of dipolar coupling observed in the spin valves, MTJs and conventional magnonic crystals.[14-17] This scenario is naturally different from a film or micron sized dots. This is an inherent property of the nanomagnets due to rebalancing of the magnetic anisotropy energy, volume energy, and surface energy. The coupling depends on the aspect ratio of the individual nanomagnets and the distance between their neighbors.

The continuous film in our experiment does not exhibit any noticeable dipolar coupling. To confirm this claim we have conducted the following experiment:

The hysteresis loops are obtained from the depth-sensitive MOKE measurement (probe fluence: 1 mJ/cm$^2$) by sweeping the external magnetic field in the out-of-plane (OOP) direction as shown in Figure S2. For the full stack, the data is obtained without the wave plate. Rotating the quarter wave plate in the probe path we have distinguished the signal from the hard magnetic layer (wave-plate angle: 50° with respect to the horizontal plane) and soft magnetic layer (wave-plate angle: 44° with respect to the horizontal plane). The coercivities for SL (CoFeB/Ta/[Tb/Co]$_5$) and RL (CoFeB) are about 1200 Oe and 100 Oe respectively. The loops are square in shape which indicates uniform reversal of the magnetic domains in the easy-axis direction (i.e., perpendicular to the sample plane). The amplitude of MOKE signal recorded from the RL is four times smaller than the SL, which is expected. A shift of about 10 Oe observed from the hysteresis loop of RL is negligible and falls within the experimental range of error.

**Different magnetization configurations during the optical switching of the MTJ cell in absence of any external magnetic field**

In Figure 2b of the paper, it is observed that almost 50% of the switching events will be successful. A few different spin configurations may form between the SL and RL in this scenario.

Here we have described the possible magnetization configurations the MTJ cell could be in, during the optical switching process:

      (i) The initial 'High' state (high resistance: magnetizations of hard and soft layer are antiparallel) is confirmed by reading the magnetoresistance in presence of a magnetic field that can essentially control the orientation of the RL (upward direction for example). Then this field is removed, and a laser pulse is launched. The SL gets switched, and the final state is read as 'Low' (low resistance: magnetizations of hard and soft layer are parallel) in presence of the field again.

      Figure 2b in the paper, shows that there are states when the switching of SL is not successful as is depicted by type (ii) in Figure S3.

      The 'Low' to 'High' state switching is also observed. It can occur in two possible ways: the SL gets switched as shown in type (iii) and RL gets reoriented by the influence of switched SL as shown in type (iv). Commonly we have observed that as soon as the top ferrimagnetic layer gets switched, the soft CoFeB layer flips its magnetization after approach of laser pulse on the MTJ cell.

      The type (v) depicts the scenario if the switching of SL is unsuccessful, and it ends up again in a 'Low' state.

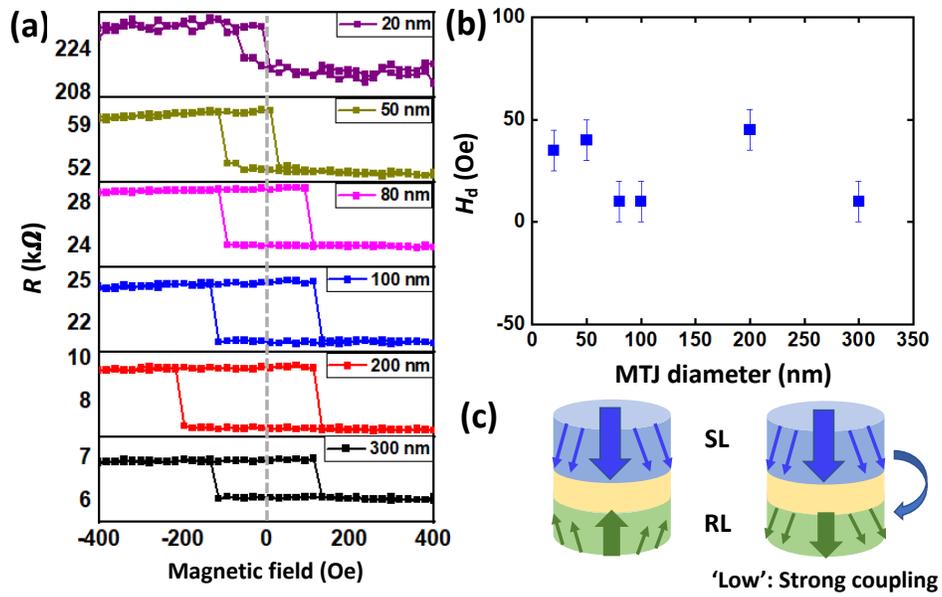

**Figure S1**: (a) The minor R-H loops obtained from MTJ cells with different diameters. The dotted line shows zero-field position. (b) The estimated dipolar field ($H_d$) plotted with cell-diameter. (c) Artist's impression of dipolar coupling between the SL and RL inside the MTJ cell.

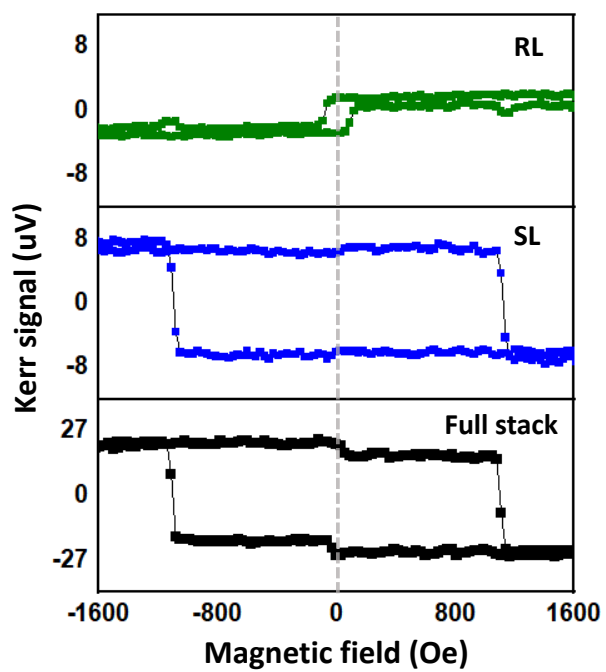

**Figure S2**: The out-of-plane hysteresis loops obtained from depth sensitive static MOKE measurements for the entire stack of the reference film, storage layer and reference layer. The dotted line represents the zero-field position.

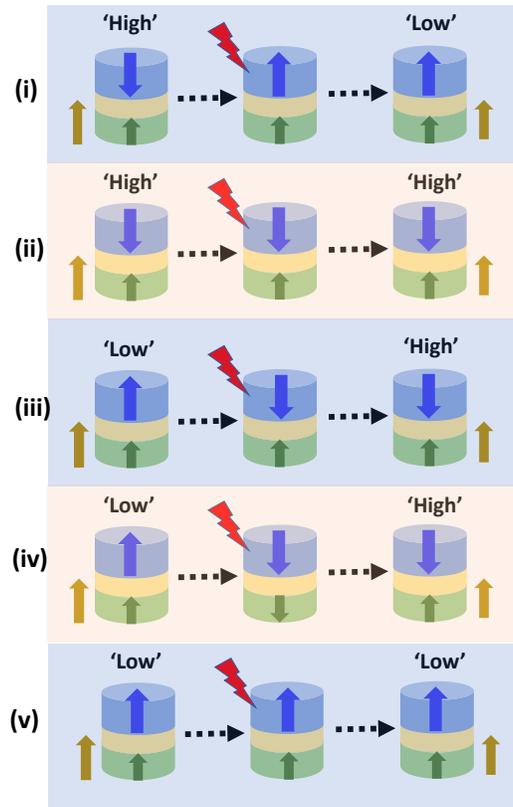

**Figure S3**: Schematic of possible magnetization configurations formed during the optical switching of the MTJ cell. The blue and red arrows represent the magnetization orientations of the hard SL and soft RL respectively. The green arrow represents the magnetic field applied during the electrical read-out operation however no field is applied during optical writing here. The drawings represent the final configuration at each stage.